\documentclass[10pt,twocolumn,prl,aps,floatfix,superscriptaddress,longbibliography]{revtex4-1}

\usepackage[utf8]{inputenc}
\usepackage{amsmath}
\usepackage{amsfonts}
\usepackage{amssymb}
\usepackage{graphicx}

\usepackage{siunitx}
\usepackage{chemformula}

\begin{document}
\title{A quantum phase transition in the one-dimensional water chain}

\author{T. Serwatka}
\affiliation{Department of Chemistry, University of Waterloo, Ontario, N2L 3G1, Canada}

\author{R. G. Melko}
\affiliation{Department of Physics \& Astronomy, University of Waterloo, Ontario, N2L 3G1, Canada}
\affiliation{Perimeter Institute for Theoretical Physics, Waterloo, Ontario N2L 2Y5, Canada}

\author{A. Burkov}
\affiliation{Department of Physics \& Astronomy, University of Waterloo, Ontario, N2L 3G1, Canada}
\affiliation{Perimeter Institute for Theoretical Physics, Waterloo, Ontario N2L 2Y5, Canada}

\author{P.-N. Roy}
\affiliation{Department of Chemistry, University of Waterloo, Ontario, N2L 3G1, Canada}
\affiliation{Perimeter Institute for Theoretical Physics, Waterloo, Ontario N2L 2Y5, Canada}

\begin{abstract}
The concept of quantum phase transitions (QPT) plays a central role in the description of condensed matter systems. In this contribution, we perform high-quality wavefunction-based simulations to demonstrate the existence of a quantum phase transition in a crucially relevant molecular system, namely water, forming linear chains of rotating molecules. We determine various critical exponents and reveal the water chain QPT to belong to the (1+1) dimensional Ising universality class. Furthermore, the effect of breaking symmetries is examined and it is shown that by breaking the inversion symmetry, the ground state degeneracy of the ordered quantum phase is lifted to yield two many-body states with opposite polarization. The possibility of forming ferroelectric phases together with a thermal stability of the quantum critical regime up to $\sim \SI{10}{\kelvin}$ makes the linear water chain a promising candidate as a platform for quantum devices.
\end{abstract}
\maketitle

Quantum phase transitions (QPT) are crucial to explain many phenomena in various condensed matter systems such as heavy-fermion compounds~\cite{si2010heavy}, magnetic materials~\cite{sachdev2012quantum}, and even unconventional superconductors~\cite{PhysRevLett.65.923,sachdev2000quantum,bianconi2001quantum}.  
{
Most recently, they have been observed in highly-controlled quantum devices \cite{samajdar2021quantum}, prepared using dipolar atoms~\cite{zhang2012observation,browaeys2016experimental} in optical lattices. Dipolar interactions between atomic or molecular~\cite{yan2013observation,hazzard2014many} species are often achieved via highly-excited electronic states, where exquisite laser and vacuum control is required to protect the species from decoherence, inter-particle collisions, chemical reactions, etc. This procedure could be avoided by using much less reactive molecules that possess a permanent dipole moment in their ground electronic state, which could be nanoconfined in a chemical environment to realize a dipolar lattice. Employing water molecules in such an endeavour would be highly desirable since water is very stable and relatively unreactive in a number of confining environments.
Moreover, water has a large permanent dipole moment, is highly abundant, and relatively inexpensive. Recent experimental results indicate quantum critical behaviour of water nanoconfined in beryl~\cite{gorshunov2016incipient,belyanchikov2022fingerprints} and cordierite~\cite{belyanchikov2020dielectric,belyanchikov2022single} crystals as well as in carbon nanotubes~\cite{PhysRevLett.118.027402}. Interestingly, in these systems water also shows ferroelectric behaviour, a property that is not known for natural, hexagonal ice. Therefore, water presents a great potential to be used to create dipolar lattices. These could eventually could be employed as quantum devices, in which the existence of quantum phase transitions would have important consequences, e.g.~for adiabatic state preparation~\cite{latorre2004adiabatic,schutzhold2006adiabatic,amin2009first} and other quantum computing algorithms. 
}

{In this study we focus on water forming linear chains as could be realized in carbon nanotubes or in partially filled crystal structures.} In this letter we predict, based on a first-principles theoretical model, a continuous QPT for a linear chain of $N$ equidistant \emph{para}-water molecules.
We use a high-quality \emph{ab initio} potential energy surface to model the water-water interactions~\cite{babin2013development,babin2014development} and employ a recently developed density matrix renormalization group approach that takes into account the rotational degrees of freedom and the nuclear spin statistics of water~\cite{serwatka2022ground}. {In all calculations we use a rather small basis set with $j_{\mathrm{max}}=1$ which turns out to be large enough to yield accurate critical exponents. The only effect of a larger basis set is a shift of the critical distance to larger distances (see supplementary material~\cite{supp}).}
Strictly speaking, a QPT only occurs in the {limit} $N\rightarrow\infty$. However, chains of finite size provide valuable precursors that already indicate the occurrence of a QPT. 
We will demonstrate by computing the excitation spectrum, entanglement measures, and correlation functions that the signs of a QPT become apparent for rather short chains containing only  {twenty or fifty water molecules (see also the supplementary material).}

\begin{figure}[h]
\centering
\includegraphics[width=\columnwidth]{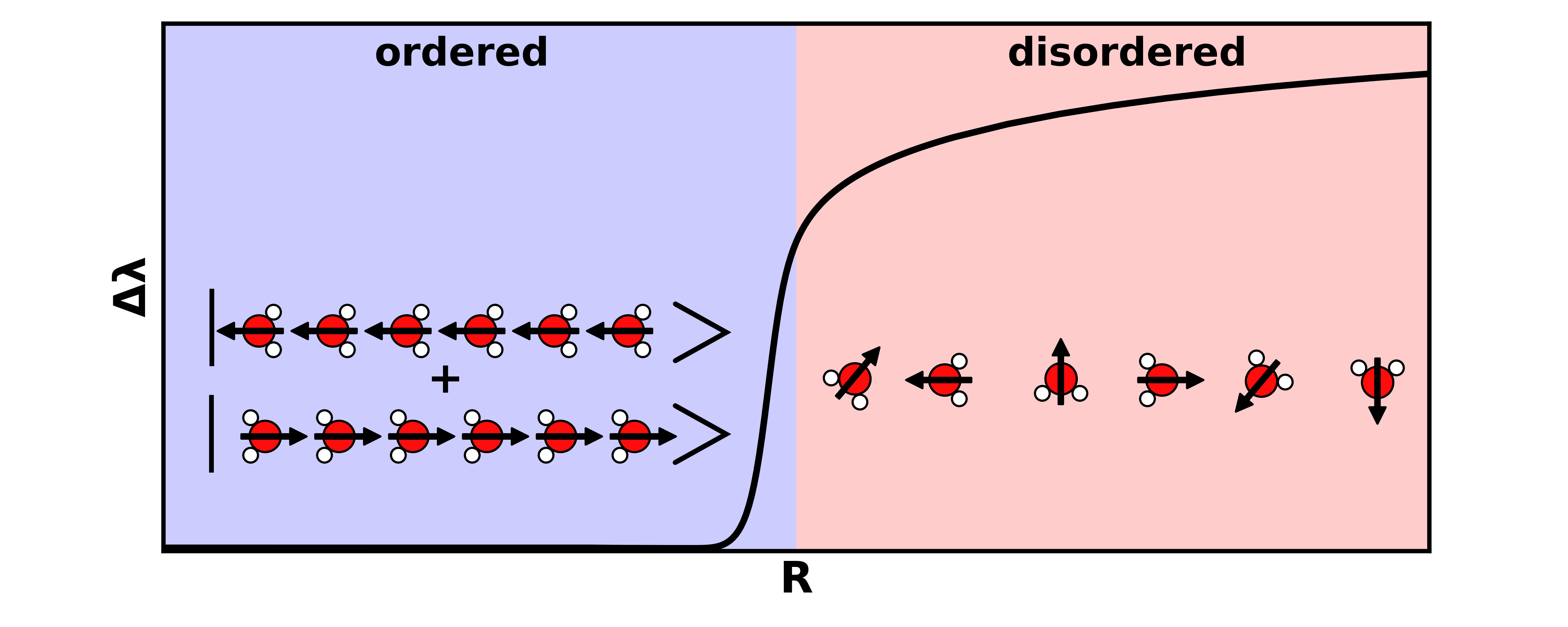}
\caption{Phase diagram of the one-dimensional water system. The disordered phase is characterized by water molecules delocalized over all orientations at large water-water distances. In contrast, in the ordered phase, the electric dipole moment (black arrows) are aligned leading to a two-fold degenerate ground state which is a superposition of a left-and right-polarized states. The solid line represents the Schmidt gap, $\Delta \lambda$, as a function of the water-water distance $R$.}
\label{fig:phase_diagram}
\end{figure}
\begin{figure}[h]
\centering
\includegraphics[width=\columnwidth]{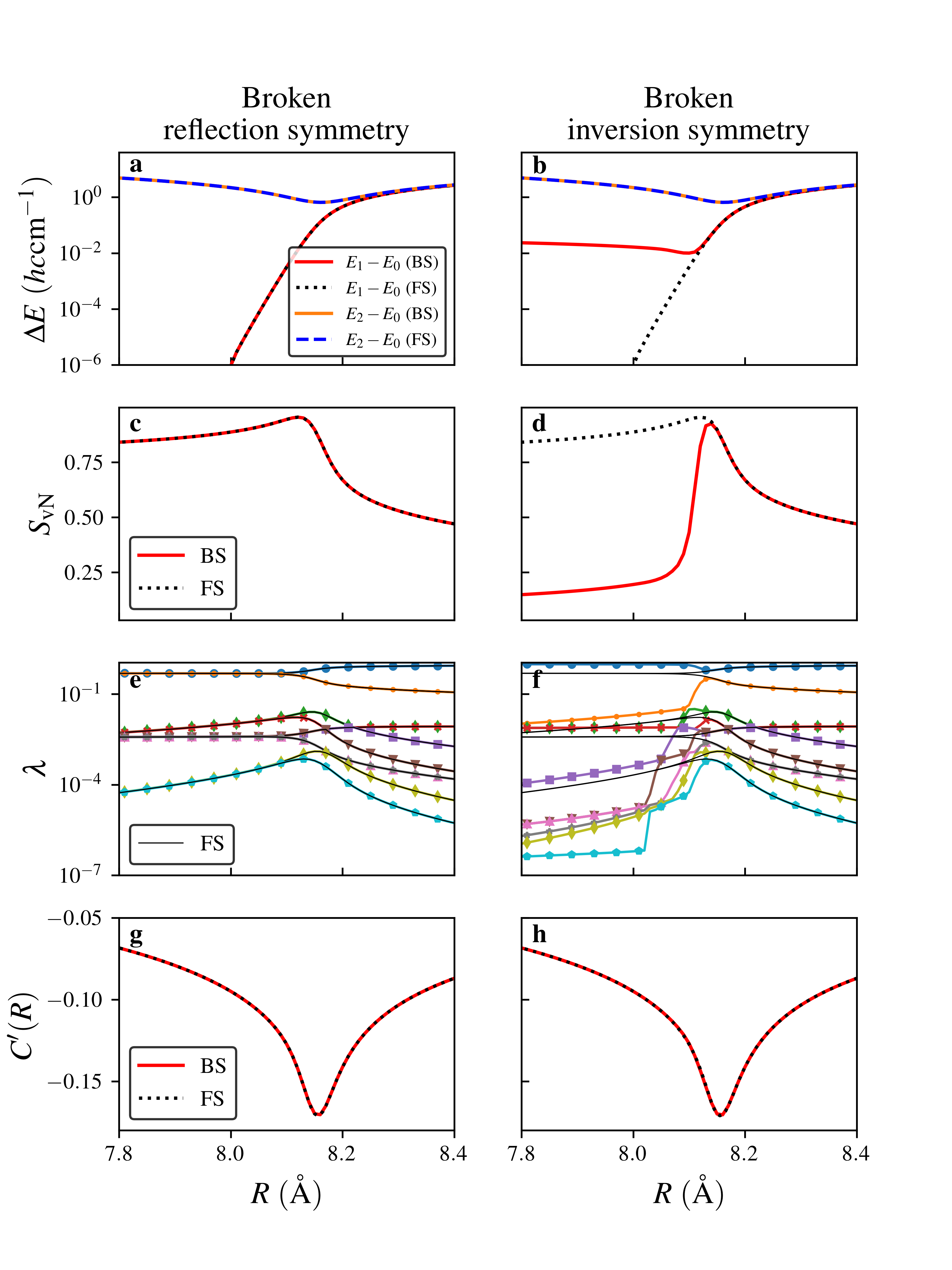}
\caption{\textbf{a}, \textbf{b} Energy difference between the ground state and the first and second excited states for chains of full symmetry (FS) and of broken symmetry (BS). \textbf{c}, \textbf{d} von-Neumann entanglement entropy for FS and BS chains. \textbf{e}, \textbf{f} First ten Schmidt values for FS and BS chains. \textbf{g}, \textbf{h} Derivative of the nearest-neighbour dipole-dipole correlation function $C(R)=\frac{1}{N-1}\sum_{i=1}^{N}\langle\mu^{z}_{i}\mu^{z}_{i+1}\rangle$. In all calculations  a small field of strength is $|\vec{E}|=\SI{e-7}{\hartree}$ applied at the two edge molecules. The system size is $N=300$.}
\label{fig:symmetry_breaking}
\end{figure}
\begin{figure}[h]
\centering
\includegraphics[width=\columnwidth]{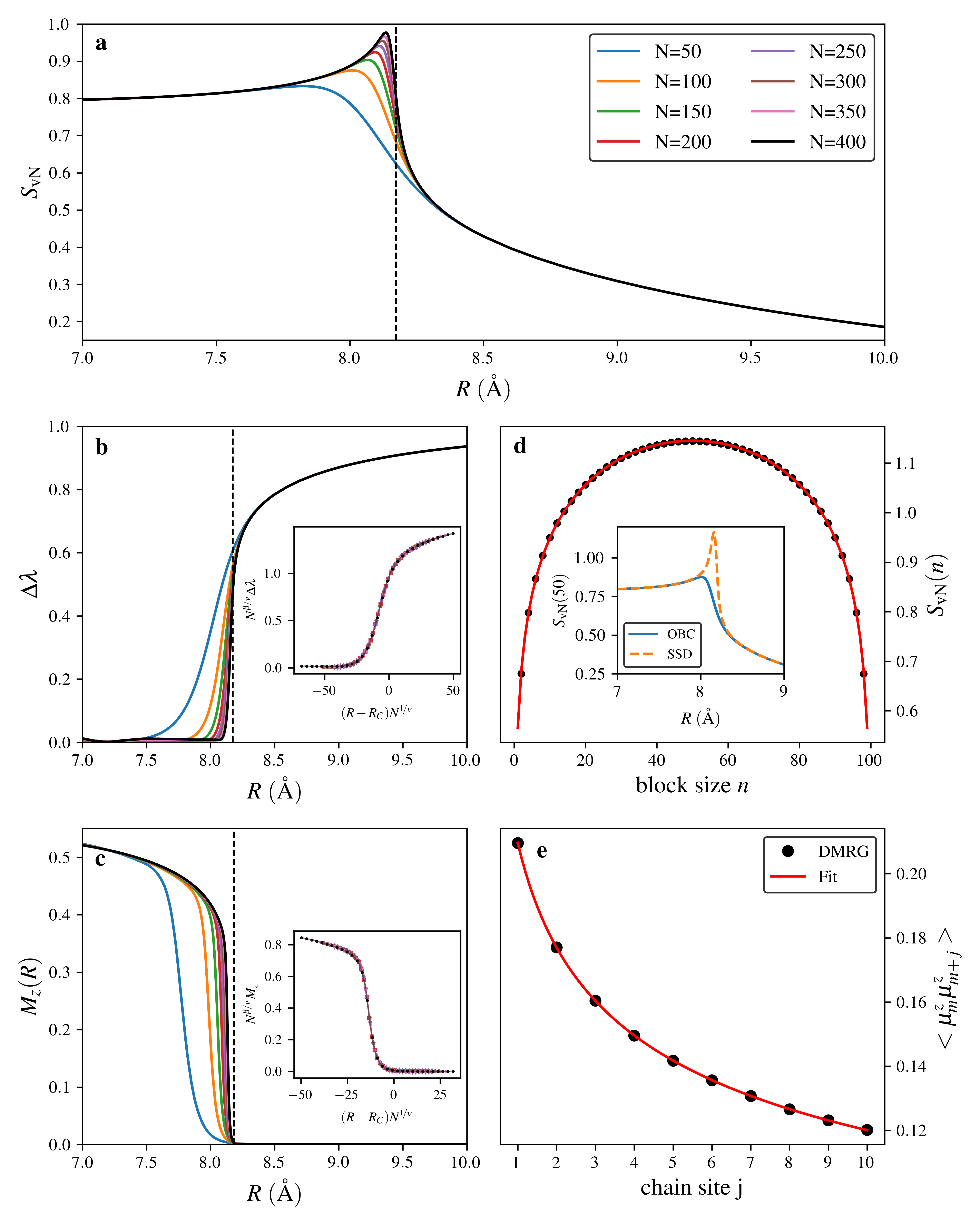}
\caption{\textbf{a} von-Neumann entanglement entropy for different numbers of water molecules. \textbf{b} Schmidt gap for different numbers of water molecules. The inset of \textbf{b} shows the rescaled Schmidt gaps (for $N=200-400$) with fitted critical exponents $\nu=1.004\pm0.0040$ and $\beta=0.118\pm 0.008$. \textbf{c} $z$-component of the total polarization per molecule for different numbers of water molecules. For these calculations the inversion symmetry was broken by applying a very small electric field ($|\vec{E}|=\SI{e-7}{\hartree}$) at the two edge molecules. The inset of \textbf{c} shows the rescaled polarization per bond (for $N=200-400$) with fitted critical exponents $\nu=1.069\pm0.025$, $\beta=0.125\pm 0.001$. The critical distance $R_{c}$ is indicated by the dashed vertical line in \textbf{a}, \textbf{b} and \textbf{c}. \textbf{d} von-Neumann entanglement entropy for different bipartitions at the QPT (solid circles). The solid line shows a fit of the entropy points to $\frac{c}{3}\ln\left(\frac{N}{\pi}\sin\left(\frac{n\pi}{N}\right)\right)$, with a central charge of $c=0.5028\pm0.0005$. The two edge points were excluded from the fit. The inset of \textbf{d} shows the von-Neumann entanglement entropy calculated with open-boundary conditions (OBC, solid line) and sine-squared deformation (SSD, dashed line). All points are calculated for $N=100$. \textbf{e} $z$-component of dipole-dipole correlation between the central water molecule at site $m=51$ and the $j$th water molecule (solid circles). The solid line shows a fit of the correlation function to $j^{-\eta}$ with a critical exponent of $\eta=0.253\pm 0.002$. All points are calculated for $N=101$.}
\label{fig:critical_coeff}
\end{figure}

 {For water at large distances $R$, interactions between the molecules} are very weak and the Hamiltonian is dominated by the rotational kinetic energy of individual molecules. The kinetic energy term tends to delocalize the ground state wave function and thus creates a phase of disordered water molecules. 
In this disordered phase, the  molecules do not adopt any specific relative orientation and can rotate nearly freely. With decreasing $R$, the dominance of the kinetic part declines because the molecular dipoles interact more strongly and start to align.  {So the water-water distance $R$ acts as a non-thermal parameter mediating the interaction strength.} Eventually, at a critical distance $R_{c}$, the system crosses the QPT and reaches a phase of ordered water molecules with aligned parallel dipoles as illustrated in Fig.~\ref{fig:phase_diagram}. The increasing interaction for decreasing distance is associated with a rise in quantum fluctuations in the ground state. At the QPT, these fluctuations diverge because the system becomes gapless and excitations with arbitrarily low energies are possible~\cite{vojta2003quantum}. 
To illustrate this situation, the energy gaps between the ground state and the first two excited states are shown in  Fig.~\ref{fig:symmetry_breaking}{\bf a} (full symmetry (FS) curves). 
After crossing the QPT, the first gap vanishes as the ground state becomes two-fold degenerate due to the inversion symmetry of the chain. 
For the second gap, and all higher excitations, we observe an avoided crossing that becomes infinitely sharp in the  {infinite-chain} limit. 
The energy scale in Fig.~\ref{fig:symmetry_breaking} also suggests that the QPT will have consequences at finite temperature $T$. Quantum fluctuations are important  {as long as they are larger or comparable to thermal fluctuations, i.e.} as long as $\hbar\omega_{c}\gtrsim k_{B}T$ where $\hbar\omega_{c}$ is the typical energy of long distance fluctuations in the order parameter~\cite{vojta2003quantum}.  {The latter is a non-thermal parameter that characterizes the ordered and disordered phase.}  {Its fluctuations are} comparable in magnitude to the fundamental energy gap. For the water chain this means that for temperature {s} up to \SI{10}{\kelvin}, there should still be a quantum-critical regime near the QPT that could be detected in experiments. For instance, a quasi-phase transition has recently been observed in single file of water molecules encapsulated in carbon nanotubes \cite{PhysRevLett.118.027402}.
When approaching the QPT, the correlations within the chain become more and more long-ranged.  {A good measure for these correlations is the von-Neumann entanglement entropy between two regions A and B of the chain
\begin{align}
S_{\mathrm{vN}}=-\mathrm{tr}\rho_{A}\mathrm{ln}\rho_{A}
\end{align}
where $\rho_{A}$ is the reduced density matrix of region A. This entanglement entropy is zero for a product state and increases with rising entanglement between region A and B.} We show the von-Neumann entanglement entropy ($S_{\mathrm{vN}}$) in Fig.~\ref{fig:critical_coeff}{\bf a}  {where we define A and B as the two halves of the chain}. At the QPT $S_{\mathrm{vN}}$ diverges because the system is in a maximally entangled state.  In this state there are quantum fluctuations on all length (and time) scales~\cite{vojta2003quantum,sachdev2011quantumcrit}. Therefore, in the critical region the correlation length is the only relevant length scale.  {Since at the QPT the correlation length $\xi$ diverges the microscopic details of the interactions become irrelevant. The system's qualitative behaviour is only determined by its symmetry and dimensionality. This leads to a universal behaviour of different systems which share the same symmetry and dimensionality. These systems can be categorized via universality classes.}
 {From the long-range correlations in the critical region it follows that} for a finite-sized system, the behaviour of the order parameter $O(R)$ (and other properties) must be unchanged if they are rescaled according to $O(R)\sim N^{\frac{\beta}{\nu}}f(|R-R_{c}|N^{\frac{1}{\nu}})$~\cite{fisher1972scaling}. 
The critical exponents $\nu$ and $\beta$ that are used to perform this scaling completely determine which universality class the system belongs to. We perform such a finite-size scaling analysis (FSSA) on the Schmidt gap, $\Delta \lambda$, as shown in Fig.~\ref{fig:critical_coeff}{\bf b}.  {The Schmidt gap is the difference between the first two eigenvalues of $\rho_{A}$.} Our analysis reveals that $\Delta \lambda$ closely follows the scaling law and is therefore an appropriate order parameter for the water chain~\cite{de2012entanglement}.
The continuous change in $\Delta \lambda$ identifies the transition as a critical point. For the critical exponents, we obtain $\nu=1.004\pm0.004$ and $\beta=0.118\pm 0.008$ which are in excellent agreement with the exact values of $\nu=1.0$ and $\beta=0.125$ expected for the (1+1)-dimensional Ising universality class~\cite{ibarra2016hobbyhorse}. 
The invariance of properties with respect to the length scale at criticality also offers the opportunity to describe such critical systems with conformal field theory (CFT)~\cite{calabrese2004entanglement,calabrese2009entanglement} via the use of a continuous quantum field description. 
In the critical region, the entanglement entropy violates the area law and is no longer bounded~\cite{eisert2010colloquium}. Instead, for a chain of $N$ sites with periodic boundary conditions CFT predicts a logarithmic scaling of the form $S\sim\frac{c}{3}\ln\left(\frac{N}{\pi}\sin\left(\frac{n\pi}{N}\right)\right)$~\cite{calabrese2004entanglement}, where $n$ is the number of sites of one of the blocks in a bipartite chain. 
The quantity $c$ is the central charge and can be determined numerically by fitting the logarithmic formula to the entanglement entropy with different bipartitions. 
The results of such an analysis is shown in Fig.~\ref{fig:critical_coeff}{\bf d} for a chain of $100$ water molecules.  
To mimic periodic boundary conditions in an open-boundary calculation we used a sine-square deformation~\cite{katsura2012sine,hotta2012grand} (SSD) to modify the Hamiltonian accordingly. 
This yields a much faster convergence of $S_{\mathrm{vN}}$ in the critical region (see inset of Fig.~\ref{fig:critical_coeff}d). By employing this procedure we obtain a central charge $c=0.5028\pm 0.0005$ which is in very good agreement with the exact value of $c=0.5$ of the $D=1$  {transverse-field Ising model} (TFIM)~\cite{calabrese2004entanglement}. 
A further critical exponent that characterizes the universality class can be obtained by considering the two-site dipole-correlation function $\langle \mu^{z}_{i}\mu^{z}_{j} \rangle$. 
In the disordered phase, the correlations remain short-ranged and the correlation function between site $i$ and $j$ falls off exponentially, i.e. $\langle \mu^{z}_{i}\mu^{z}_{j} \rangle \sim e^{-\frac{|i-j|}{\xi}}$ for large $|i-j|$. 
At the QPT, due to the diverging correlation length $\xi$, the correlation function follows a power law, i.e. $\langle \mu^{z}_{i}\mu^{z}_{j} \rangle \sim \frac{1}{|i-j|^{\eta}}$ uniquely determined by the critical exponent $\eta$, the so-called anomalous dimension. 
Fitting the $z$-component of the electric dipole-dipole correlation between the central water molecule and a water molecule at site $j$ to such a power law yields a critical coefficient $\eta=0.253\pm 0.002$ (see Fig.~\ref{fig:critical_coeff}e) that is again in excellent agreement with the (1+1)-dimensional Ising universality class value ($\eta=0.25$~\cite{ibarra2016hobbyhorse}). The fact that the QPT in the water chains belongs to the (1+1)-dimensional Ising universality class can be understood via inspection of the respective Hamiltonians. Both systems have on-site one-body terms, the kinetic energy for the water chain, and the transverse field for the TFIM, and each model also has two-body interaction terms along one spatial axis.

\begin{figure}[h]
\centering
\includegraphics[width=\columnwidth]{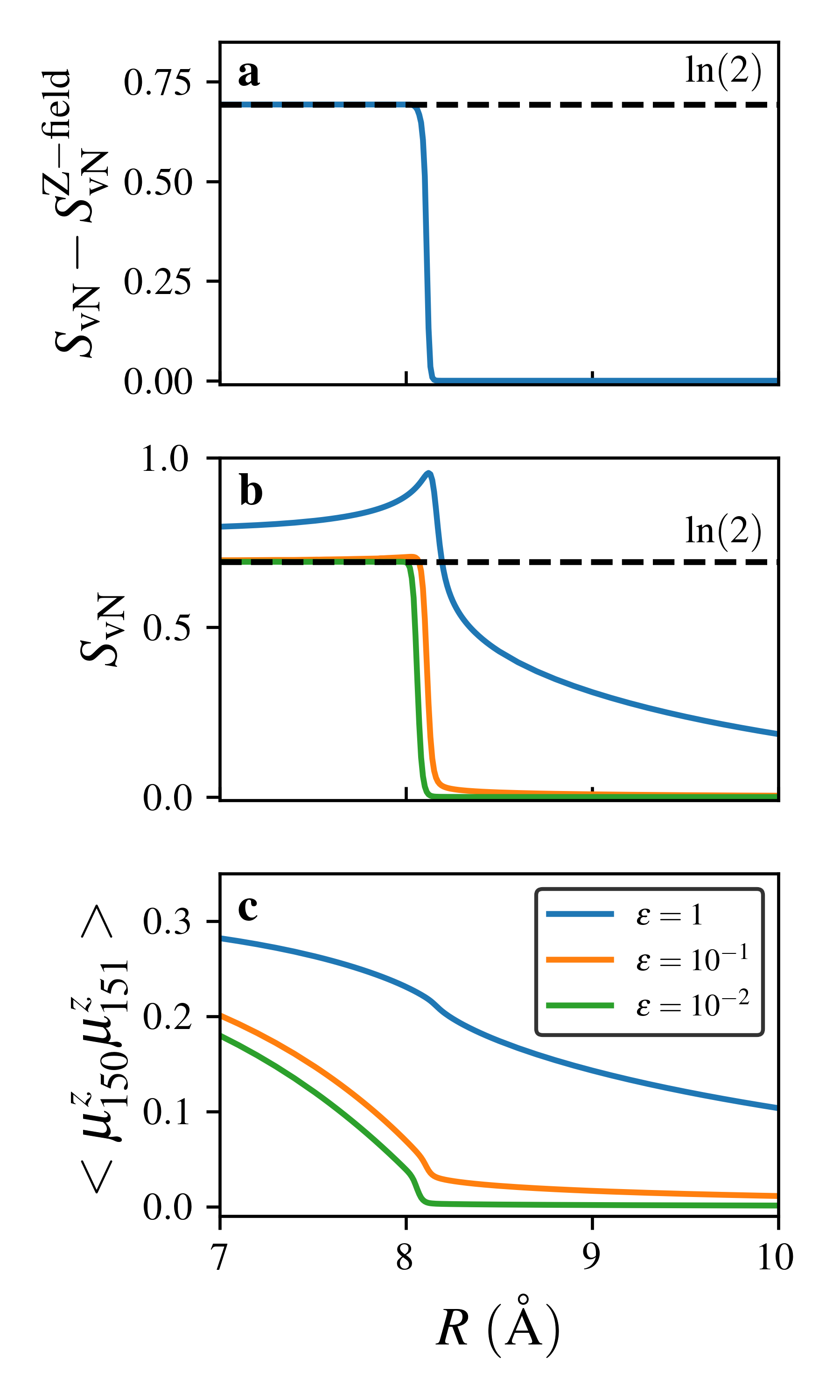}
\caption{\textbf{a} Difference of the von-Neumann entanglement entropy of a fully symmetric chain and a chain with broken inversion symmetry. \textbf{b} von-Neumann entanglement entropy of a fully symmetric chain with a central bond weakend by a factor $\varepsilon$. The dashed line indicates the value of $\ln\left(2\right)$. \textbf{c} $z$-component of the electric dipole-dipole correlation of the two central water molecules of a fully symmetric chain with a central bond weakend by a factor $\varepsilon$. The system size is $N=300$.}
\label{fig:bond_weakening}
\end{figure}

Finally, the effect of breaking certain symmetries is considered.
To break chosen symmetries, a very small electric field is introduced at the two edge water molecules. Two cases are investigated: a transverse $XY$-polarized field to break the reflection symmetry in the planes containing the chain axis and a $Z$-polarized longitudinal field to break the inversion symmetry of the chain. We present in Fig.~\ref{fig:symmetry_breaking} the effects of symmetry breaking on a number of properties. 
We first observe that breaking the reflection symmetry has no effect on the QPT or quantum phases. 
In contrast, breaking  inversion symmetry changes the ordered quantum phase fundamentally. 
It lifts the degeneracy of the ground state, an effect similar to the spontaneous symmetry breaking in the TFIM. The equally-weighted superposition of a left-and right-polarized state (see Fig.~\ref{fig:phase_diagram}) splits up in two states with opposite polarization. 
By lifting the degeneracy, the major part of the entanglement entropy is removed (see Fig.~\ref{fig:symmetry_breaking}{\bf d}). This becomes even more obvious if one looks at the entropy difference of the fully symmetric chain and the system with broken inversion symmetry in Fig.~\ref{fig:bond_weakening}{\bf a}.
We observe a large entropy plateau after the QPT  due to the two-fold degeneracy of the ground state. This two-fold degeneracy in the fully-symmetric chain also leads to the effect that in a bipartite chain, which is in the ordered phase, the interaction strength between the two halves can be made infinitesimally small without removing the entanglement between the halves. 
This is shown in Fig.~\ref{fig:bond_weakening}{\bf b} where the central bond was weakened successively and the entanglement entropy in the ordered phase goes to $\ln (2)$, exactly the value of a two-fold degenerate state. 
In practice, this pure quantum effect implies that even if one could weaken the central bond substantially, e.g. by removing a central water molecule or by inserting fullerene cages, the two halves would still be entangled. 
This means that the electric dipole moments of the molecules located to the left and to the right of the weakened bond are still correlated (see Fig.~\ref{fig:bond_weakening}{\bf c}). 
This entanglement might be of practical use in the context of quantum information processing.
The change in entanglement that results from breaking inversion symmetry also affects the  {eigenvalues of $\rho_{A}$},  {also known as} Schmidt coefficients (see Fig.~\ref{fig:symmetry_breaking}{\bf f}). 
The Schmidt gap no longer acts as an order parameter and the whole Schmidt spectrum changes not only quantitatively but qualitatively. 
In the fully symmetric chain, all Schmidt values occur in multiples of two. This degeneracy pattern vanishes if the inversion symmetry is broken. Interestingly, in contrast to entanglement entropy and Schmidt spectrum, the  {derivative of the} nearest-neighbour correlation functions as well as the energy gap to higher-lying excited states  {are} not affected by removing the inversion symmetry. They still signal the occurrence of a QPT. And indeed, if one performs a FSSA at the $z$-component of the total polarization per molecule that now becomes a proper order parameter, the critical exponents of the (1+1)-dimensional Ising universality class are obtained. Hence, neither the QPT nor the universality class is altered by breaking the inversion symmetry. It is only the nature of the ordered phase that is affected. \\

In future work, this nature should be studied more deeply since understanding the system's entanglement properties might lead to the application of linear water chains as quantum information processing devices~\cite{halverson2018quantifying,biskupek2020bond} for which a rich landscape of schemes exists for the quantum control and information encoding in molecular rotations~\cite{koch2019quantum,albert2020robust,hughes2020robust,tscherbul2022robust}.  {Water systems in different chemical environments will be studied especially with regard to} promising experimental realizations of confined water with intermolecular distances in the quantum critical regime which have recently been achieved in beryl~\cite{gorshunov2016incipient} and cordierite~\cite{belyanchikov2020dielectric} with fascinating ferroelectric behaviour.\\

\begin{acknowledgments}
This research was supported by the Natural Sciences and Engineering Research Council of Canada (NSERC) (RGPIN-2016-04403), the Ontario Ministry of Research and Innovation (MRI), the Canada Research Chair program (950-231024), the Canada Foundation for Innovation (CFI) (project No. 35232), Compute Canada, and the Perimeter Institute for Theoretical Physics. Research at Perimeter Institute is supported in part by the Government of Canada through the Department of Innovation, Science and Economic Development Canada and by the Province of Ontario through the Ministry of Colleges and Universities.
\end{acknowledgments}


\begin{thebibliography}{41}%
\makeatletter
\providecommand \@ifxundefined [1]{%
 \@ifx{#1\undefined}
}%
\providecommand \@ifnum [1]{%
 \ifnum #1\expandafter \@firstoftwo
 \else \expandafter \@secondoftwo
 \fi
}%
\providecommand \@ifx [1]{%
 \ifx #1\expandafter \@firstoftwo
 \else \expandafter \@secondoftwo
 \fi
}%
\providecommand \natexlab [1]{#1}%
\providecommand \enquote  [1]{``#1''}%
\providecommand \bibnamefont  [1]{#1}%
\providecommand \bibfnamefont [1]{#1}%
\providecommand \citenamefont [1]{#1}%
\providecommand \href@noop [0]{\@secondoftwo}%
\providecommand \href [0]{\begingroup \@sanitize@url \@href}%
\providecommand \@href[1]{\@@startlink{#1}\@@href}%
\providecommand \@@href[1]{\endgroup#1\@@endlink}%
\providecommand \@sanitize@url [0]{\catcode `\\12\catcode `\$12\catcode
  `\&12\catcode `\#12\catcode `\^12\catcode `\_12\catcode `\%12\relax}%
\providecommand \@@startlink[1]{}%
\providecommand \@@endlink[0]{}%
\providecommand \url  [0]{\begingroup\@sanitize@url \@url }%
\providecommand \@url [1]{\endgroup\@href {#1}{\urlprefix }}%
\providecommand \urlprefix  [0]{URL }%
\providecommand \Eprint [0]{\href }%
\providecommand \doibase [0]{http://dx.doi.org/}%
\providecommand \selectlanguage [0]{\@gobble}%
\providecommand \bibinfo  [0]{\@secondoftwo}%
\providecommand \bibfield  [0]{\@secondoftwo}%
\providecommand \translation [1]{[#1]}%
\providecommand \BibitemOpen [0]{}%
\providecommand \bibitemStop [0]{}%
\providecommand \bibitemNoStop [0]{.\EOS\space}%
\providecommand \EOS [0]{\spacefactor3000\relax}%
\providecommand \BibitemShut  [1]{\csname bibitem#1\endcsname}%
\let\auto@bib@innerbib\@empty
\bibitem [{\citenamefont {Si}\ and\ \citenamefont
  {Steglich}(2010)}]{si2010heavy}%
  \BibitemOpen
  \bibfield  {author} {\bibinfo {author} {\bibfnamefont {Qimiao}\ \bibnamefont
  {Si}}\ and\ \bibinfo {author} {\bibfnamefont {Frank}\ \bibnamefont
  {Steglich}},\ }\bibfield  {title} {\enquote {\bibinfo {title} {Heavy fermions
  and quantum phase transitions},}\ }\href@noop {} {\bibfield  {journal}
  {\bibinfo  {journal} {Science}\ }\textbf {\bibinfo {volume} {329}},\ \bibinfo
  {pages} {1161--1166} (\bibinfo {year} {2010})}\BibitemShut {NoStop}%
\bibitem [{\citenamefont {Sachdev}(2012)}]{sachdev2012quantum}%
  \BibitemOpen
  \bibfield  {author} {\bibinfo {author} {\bibfnamefont {Subir}\ \bibnamefont
  {Sachdev}},\ }\bibfield  {title} {\enquote {\bibinfo {title} {Quantum phase
  transitions of antiferromagnets and the cuprate superconductors},}\ }in\
  \href@noop {} {\emph {\bibinfo {booktitle} {Modern theories of many-particle
  systems in condensed matter physics}}}\ (\bibinfo  {publisher} {Springer},\
  \bibinfo {year} {2012})\ pp.\ \bibinfo {pages} {1--51}\BibitemShut {NoStop}%
\bibitem [{\citenamefont {Fisher}(1990)}]{PhysRevLett.65.923}%
  \BibitemOpen
  \bibfield  {author} {\bibinfo {author} {\bibfnamefont {Matthew P.~A.}\
  \bibnamefont {Fisher}},\ }\bibfield  {title} {\enquote {\bibinfo {title}
  {Quantum phase transitions in disordered two-dimensional superconductors},}\
  }\href {\doibase 10.1103/PhysRevLett.65.923} {\bibfield  {journal} {\bibinfo
  {journal} {Phys. Rev. Lett.}\ }\textbf {\bibinfo {volume} {65}},\ \bibinfo
  {pages} {923--926} (\bibinfo {year} {1990})}\BibitemShut {NoStop}%
\bibitem [{\citenamefont {Sachdev}(2000)}]{sachdev2000quantum}%
  \BibitemOpen
  \bibfield  {author} {\bibinfo {author} {\bibfnamefont {Subir}\ \bibnamefont
  {Sachdev}},\ }\bibfield  {title} {\enquote {\bibinfo {title} {Quantum
  criticality: competing ground states in low dimensions},}\ }\href@noop {}
  {\bibfield  {journal} {\bibinfo  {journal} {Science}\ }\textbf {\bibinfo
  {volume} {288}},\ \bibinfo {pages} {475--480} (\bibinfo {year}
  {2000})}\BibitemShut {NoStop}%
\bibitem [{\citenamefont {Bianconi}\ \emph {et~al.}(2001)\citenamefont
  {Bianconi}, \citenamefont {Agrestini}, \citenamefont {Bianconi},
  \citenamefont {Di~Castro},\ and\ \citenamefont
  {Saini}}]{bianconi2001quantum}%
  \BibitemOpen
  \bibfield  {author} {\bibinfo {author} {\bibfnamefont {A.}~\bibnamefont
  {Bianconi}}, \bibinfo {author} {\bibfnamefont {S.}~\bibnamefont {Agrestini}},
  \bibinfo {author} {\bibfnamefont {G.}~\bibnamefont {Bianconi}}, \bibinfo
  {author} {\bibfnamefont {D.}~\bibnamefont {Di~Castro}}, \ and\ \bibinfo
  {author} {\bibfnamefont {NL}~\bibnamefont {Saini}},\ }\bibfield  {title}
  {\enquote {\bibinfo {title} {A quantum phase transition driven by the
  electron lattice interaction gives high tc superconductivity},}\ }\href@noop
  {} {\bibfield  {journal} {\bibinfo  {journal} {J. Alloys Compd.}\ }\textbf
  {\bibinfo {volume} {317}},\ \bibinfo {pages} {537--541} (\bibinfo {year}
  {2001})}\BibitemShut {NoStop}%
\bibitem [{\citenamefont {Samajdar}\ \emph {et~al.}(2021)\citenamefont
  {Samajdar}, \citenamefont {Ho}, \citenamefont {Pichler}, \citenamefont
  {Lukin},\ and\ \citenamefont {Sachdev}}]{samajdar2021quantum}%
  \BibitemOpen
  \bibfield  {author} {\bibinfo {author} {\bibfnamefont {Rhine}\ \bibnamefont
  {Samajdar}}, \bibinfo {author} {\bibfnamefont {Wen~Wei}\ \bibnamefont {Ho}},
  \bibinfo {author} {\bibfnamefont {Hannes}\ \bibnamefont {Pichler}}, \bibinfo
  {author} {\bibfnamefont {Mikhail~D}\ \bibnamefont {Lukin}}, \ and\ \bibinfo
  {author} {\bibfnamefont {Subir}\ \bibnamefont {Sachdev}},\ }\bibfield
  {title} {\enquote {\bibinfo {title} {Quantum phases of rydberg atoms on a
  kagome lattice},}\ }\href@noop {} {\bibfield  {journal} {\bibinfo  {journal}
  {Proceedings of the National Academy of Sciences}\ }\textbf {\bibinfo
  {volume} {118}} (\bibinfo {year} {2021})}\BibitemShut {NoStop}%
\bibitem [{\citenamefont {Zhang}\ \emph {et~al.}(2012)\citenamefont {Zhang},
  \citenamefont {Hung}, \citenamefont {Tung},\ and\ \citenamefont
  {Chin}}]{zhang2012observation}%
  \BibitemOpen
  \bibfield  {author} {\bibinfo {author} {\bibfnamefont {Xibo}\ \bibnamefont
  {Zhang}}, \bibinfo {author} {\bibfnamefont {Chen-Lung}\ \bibnamefont {Hung}},
  \bibinfo {author} {\bibfnamefont {Shih-Kuang}\ \bibnamefont {Tung}}, \ and\
  \bibinfo {author} {\bibfnamefont {Cheng}\ \bibnamefont {Chin}},\ }\bibfield
  {title} {\enquote {\bibinfo {title} {Observation of quantum criticality with
  ultracold atoms in optical lattices},}\ }\href@noop {} {\bibfield  {journal}
  {\bibinfo  {journal} {Science}\ }\textbf {\bibinfo {volume} {335}},\ \bibinfo
  {pages} {1070--1072} (\bibinfo {year} {2012})}\BibitemShut {NoStop}%
\bibitem [{\citenamefont {Browaeys}\ \emph {et~al.}(2016)\citenamefont
  {Browaeys}, \citenamefont {Barredo},\ and\ \citenamefont
  {Lahaye}}]{browaeys2016experimental}%
  \BibitemOpen
  \bibfield  {author} {\bibinfo {author} {\bibfnamefont {Antoine}\ \bibnamefont
  {Browaeys}}, \bibinfo {author} {\bibfnamefont {Daniel}\ \bibnamefont
  {Barredo}}, \ and\ \bibinfo {author} {\bibfnamefont {Thierry}\ \bibnamefont
  {Lahaye}},\ }\bibfield  {title} {\enquote {\bibinfo {title} {Experimental
  investigations of dipole--dipole interactions between a few rydberg atoms},}\
  }\href@noop {} {\bibfield  {journal} {\bibinfo  {journal} {J. Phys. B: At.
  Mol. Opt. Phys.}\ }\textbf {\bibinfo {volume} {49}},\ \bibinfo {pages}
  {152001} (\bibinfo {year} {2016})}\BibitemShut {NoStop}%
\bibitem [{\citenamefont {Yan}\ \emph {et~al.}(2013)\citenamefont {Yan},
  \citenamefont {Moses}, \citenamefont {Gadway}, \citenamefont {Covey},
  \citenamefont {Hazzard}, \citenamefont {Rey}, \citenamefont {Jin},\ and\
  \citenamefont {Ye}}]{yan2013observation}%
  \BibitemOpen
  \bibfield  {author} {\bibinfo {author} {\bibfnamefont {Bo}~\bibnamefont
  {Yan}}, \bibinfo {author} {\bibfnamefont {Steven~A}\ \bibnamefont {Moses}},
  \bibinfo {author} {\bibfnamefont {Bryce}\ \bibnamefont {Gadway}}, \bibinfo
  {author} {\bibfnamefont {Jacob~P}\ \bibnamefont {Covey}}, \bibinfo {author}
  {\bibfnamefont {Kaden~RA}\ \bibnamefont {Hazzard}}, \bibinfo {author}
  {\bibfnamefont {Ana~Maria}\ \bibnamefont {Rey}}, \bibinfo {author}
  {\bibfnamefont {Deborah~S}\ \bibnamefont {Jin}}, \ and\ \bibinfo {author}
  {\bibfnamefont {Jun}\ \bibnamefont {Ye}},\ }\bibfield  {title} {\enquote
  {\bibinfo {title} {Observation of dipolar spin-exchange interactions with
  lattice-confined polar molecules},}\ }\href@noop {} {\bibfield  {journal}
  {\bibinfo  {journal} {Nature}\ }\textbf {\bibinfo {volume} {501}},\ \bibinfo
  {pages} {521--525} (\bibinfo {year} {2013})}\BibitemShut {NoStop}%
\bibitem [{\citenamefont {Hazzard}\ \emph {et~al.}(2014)\citenamefont
  {Hazzard}, \citenamefont {Gadway}, \citenamefont {Foss-Feig}, \citenamefont
  {Yan}, \citenamefont {Moses}, \citenamefont {Covey}, \citenamefont {Yao},
  \citenamefont {Lukin}, \citenamefont {Ye}, \citenamefont {Jin},\ and\
  \citenamefont {Rey}}]{hazzard2014many}%
  \BibitemOpen
  \bibfield  {author} {\bibinfo {author} {\bibfnamefont {Kaden R.~A.}\
  \bibnamefont {Hazzard}}, \bibinfo {author} {\bibfnamefont {Bryce}\
  \bibnamefont {Gadway}}, \bibinfo {author} {\bibfnamefont {Michael}\
  \bibnamefont {Foss-Feig}}, \bibinfo {author} {\bibfnamefont {Bo}~\bibnamefont
  {Yan}}, \bibinfo {author} {\bibfnamefont {Steven~A.}\ \bibnamefont {Moses}},
  \bibinfo {author} {\bibfnamefont {Jacob~P.}\ \bibnamefont {Covey}}, \bibinfo
  {author} {\bibfnamefont {Norman~Y.}\ \bibnamefont {Yao}}, \bibinfo {author}
  {\bibfnamefont {Mikhail~D.}\ \bibnamefont {Lukin}}, \bibinfo {author}
  {\bibfnamefont {Jun}\ \bibnamefont {Ye}}, \bibinfo {author} {\bibfnamefont
  {Deborah~S.}\ \bibnamefont {Jin}}, \ and\ \bibinfo {author} {\bibfnamefont
  {Ana~Maria}\ \bibnamefont {Rey}},\ }\bibfield  {title} {\enquote {\bibinfo
  {title} {Many-body dynamics of dipolar molecules in an optical lattice},}\
  }\href {\doibase 10.1103/PhysRevLett.113.195302} {\bibfield  {journal}
  {\bibinfo  {journal} {Phys. Rev. Lett.}\ }\textbf {\bibinfo {volume} {113}},\
  \bibinfo {pages} {195302} (\bibinfo {year} {2014})}\BibitemShut {NoStop}%
\bibitem [{\citenamefont {Gorshunov}\ \emph {et~al.}(2016)\citenamefont
  {Gorshunov}, \citenamefont {Torgashev}, \citenamefont {Zhukova},
  \citenamefont {Thomas}, \citenamefont {Belyanchikov}, \citenamefont {Kadlec},
  \citenamefont {Kadlec}, \citenamefont {Savinov}, \citenamefont {Ostapchuk},
  \citenamefont {Petzelt} \emph {et~al.}}]{gorshunov2016incipient}%
  \BibitemOpen
  \bibfield  {author} {\bibinfo {author} {\bibfnamefont {BP}~\bibnamefont
  {Gorshunov}}, \bibinfo {author} {\bibfnamefont {VI}~\bibnamefont
  {Torgashev}}, \bibinfo {author} {\bibfnamefont {ES}~\bibnamefont {Zhukova}},
  \bibinfo {author} {\bibfnamefont {VG}~\bibnamefont {Thomas}}, \bibinfo
  {author} {\bibfnamefont {MA}~\bibnamefont {Belyanchikov}}, \bibinfo {author}
  {\bibfnamefont {C}~\bibnamefont {Kadlec}}, \bibinfo {author} {\bibfnamefont
  {F}~\bibnamefont {Kadlec}}, \bibinfo {author} {\bibfnamefont {M}~\bibnamefont
  {Savinov}}, \bibinfo {author} {\bibfnamefont {T}~\bibnamefont {Ostapchuk}},
  \bibinfo {author} {\bibfnamefont {J}~\bibnamefont {Petzelt}},  \emph
  {et~al.},\ }\bibfield  {title} {\enquote {\bibinfo {title} {Incipient
  ferroelectricity of water molecules confined to nano-channels of beryl},}\
  }\href@noop {} {\bibfield  {journal} {\bibinfo  {journal} {Nature
  communications}\ }\textbf {\bibinfo {volume} {7}},\ \bibinfo {pages} {1--10}
  (\bibinfo {year} {2016})}\BibitemShut {NoStop}%
\bibitem [{\citenamefont {Belyanchikov}\ \emph
  {et~al.}(2022{\natexlab{a}})\citenamefont {Belyanchikov}, \citenamefont
  {Savinov}, \citenamefont {Proschek}, \citenamefont {Prokleska}, \citenamefont
  {Zhukova}, \citenamefont {Thomas}, \citenamefont {Bedran}, \citenamefont
  {Kadlec}, \citenamefont {Kamba}, \citenamefont {Dressel} \emph
  {et~al.}}]{belyanchikov2022fingerprints}%
  \BibitemOpen
  \bibfield  {author} {\bibinfo {author} {\bibfnamefont {Mikhail~A}\
  \bibnamefont {Belyanchikov}}, \bibinfo {author} {\bibfnamefont {Maxim}\
  \bibnamefont {Savinov}}, \bibinfo {author} {\bibfnamefont {Petr}\
  \bibnamefont {Proschek}}, \bibinfo {author} {\bibfnamefont {Jan}\
  \bibnamefont {Prokleska}}, \bibinfo {author} {\bibfnamefont {Elena~S}\
  \bibnamefont {Zhukova}}, \bibinfo {author} {\bibfnamefont {Victor~G}\
  \bibnamefont {Thomas}}, \bibinfo {author} {\bibfnamefont {Zakhar~V}\
  \bibnamefont {Bedran}}, \bibinfo {author} {\bibfnamefont {Filip}\
  \bibnamefont {Kadlec}}, \bibinfo {author} {\bibfnamefont {Stanislav}\
  \bibnamefont {Kamba}}, \bibinfo {author} {\bibfnamefont {Martin}\
  \bibnamefont {Dressel}},  \emph {et~al.},\ }\bibfield  {title} {\enquote
  {\bibinfo {title} {Fingerprints of critical phenomena in a quantum
  paraelectric ensemble of nanoconfined water molecules},}\ }\href@noop {}
  {\bibfield  {journal} {\bibinfo  {journal} {Nano Lett.}\ }\textbf {\bibinfo
  {volume} {22}},\ \bibinfo {pages} {3380--3384} (\bibinfo {year}
  {2022}{\natexlab{a}})}\BibitemShut {NoStop}%
\bibitem [{\citenamefont {Belyanchikov}\ \emph {et~al.}(2020)\citenamefont
  {Belyanchikov}, \citenamefont {Savinov}, \citenamefont {Bedran},
  \citenamefont {Bednyakov}, \citenamefont {Proschek}, \citenamefont
  {Prokleska}, \citenamefont {Abalmasov}, \citenamefont {Petzelt},
  \citenamefont {Zhukova}, \citenamefont {Thomas} \emph
  {et~al.}}]{belyanchikov2020dielectric}%
  \BibitemOpen
  \bibfield  {author} {\bibinfo {author} {\bibfnamefont {MA}~\bibnamefont
  {Belyanchikov}}, \bibinfo {author} {\bibfnamefont {M}~\bibnamefont
  {Savinov}}, \bibinfo {author} {\bibfnamefont {ZV}~\bibnamefont {Bedran}},
  \bibinfo {author} {\bibfnamefont {P}~\bibnamefont {Bednyakov}}, \bibinfo
  {author} {\bibfnamefont {P}~\bibnamefont {Proschek}}, \bibinfo {author}
  {\bibfnamefont {J}~\bibnamefont {Prokleska}}, \bibinfo {author}
  {\bibfnamefont {VA}~\bibnamefont {Abalmasov}}, \bibinfo {author}
  {\bibfnamefont {J}~\bibnamefont {Petzelt}}, \bibinfo {author} {\bibfnamefont
  {ES}~\bibnamefont {Zhukova}}, \bibinfo {author} {\bibfnamefont
  {VG}~\bibnamefont {Thomas}},  \emph {et~al.},\ }\bibfield  {title} {\enquote
  {\bibinfo {title} {Dielectric ordering of water molecules arranged in a
  dipolar lattice},}\ }\href@noop {} {\bibfield  {journal} {\bibinfo  {journal}
  {Nat. Commun.}\ }\textbf {\bibinfo {volume} {11}},\ \bibinfo {pages} {1--9}
  (\bibinfo {year} {2020})}\BibitemShut {NoStop}%
\bibitem [{\citenamefont {Belyanchikov}\ \emph
  {et~al.}(2022{\natexlab{b}})\citenamefont {Belyanchikov}, \citenamefont
  {Bedran}, \citenamefont {Savinov}, \citenamefont {Bednyakov}, \citenamefont
  {Proschek}, \citenamefont {Prokleska}, \citenamefont {Abalmasov},
  \citenamefont {Zhukova}, \citenamefont {Thomas}, \citenamefont {Dudka} \emph
  {et~al.}}]{belyanchikov2022single}%
  \BibitemOpen
  \bibfield  {author} {\bibinfo {author} {\bibfnamefont {MA}~\bibnamefont
  {Belyanchikov}}, \bibinfo {author} {\bibfnamefont {ZV}~\bibnamefont
  {Bedran}}, \bibinfo {author} {\bibfnamefont {M}~\bibnamefont {Savinov}},
  \bibinfo {author} {\bibfnamefont {P}~\bibnamefont {Bednyakov}}, \bibinfo
  {author} {\bibfnamefont {P}~\bibnamefont {Proschek}}, \bibinfo {author}
  {\bibfnamefont {J}~\bibnamefont {Prokleska}}, \bibinfo {author}
  {\bibfnamefont {VA}~\bibnamefont {Abalmasov}}, \bibinfo {author}
  {\bibfnamefont {ES}~\bibnamefont {Zhukova}}, \bibinfo {author} {\bibfnamefont
  {VG}~\bibnamefont {Thomas}}, \bibinfo {author} {\bibfnamefont
  {A}~\bibnamefont {Dudka}},  \emph {et~al.},\ }\bibfield  {title} {\enquote
  {\bibinfo {title} {Single-particle and collective excitations of polar water
  molecules confined in nano-pores within a cordierite crystal lattice},}\
  }\href@noop {} {\bibfield  {journal} {\bibinfo  {journal} {Phys. Chem. Chem.
  Phys.}\ }\textbf {\bibinfo {volume} {24}},\ \bibinfo {pages} {6890--6904}
  (\bibinfo {year} {2022}{\natexlab{b}})}\BibitemShut {NoStop}%
\bibitem [{\citenamefont {Ma}\ \emph {et~al.}(2017)\citenamefont {Ma},
  \citenamefont {Cambr\'e}, \citenamefont {Wenseleers}, \citenamefont {Doorn},\
  and\ \citenamefont {Htoon}}]{PhysRevLett.118.027402}%
  \BibitemOpen
  \bibfield  {author} {\bibinfo {author} {\bibfnamefont {Xuedan}\ \bibnamefont
  {Ma}}, \bibinfo {author} {\bibfnamefont {Sofie}\ \bibnamefont {Cambr\'e}},
  \bibinfo {author} {\bibfnamefont {Wim}\ \bibnamefont {Wenseleers}}, \bibinfo
  {author} {\bibfnamefont {Stephen~K.}\ \bibnamefont {Doorn}}, \ and\ \bibinfo
  {author} {\bibfnamefont {Han}\ \bibnamefont {Htoon}},\ }\bibfield  {title}
  {\enquote {\bibinfo {title} {Quasiphase transition in a single file of water
  molecules encapsulated in (6,5) carbon nanotubes observed by
  temperature-dependent photoluminescence spectroscopy},}\ }\href {\doibase
  10.1103/PhysRevLett.118.027402} {\bibfield  {journal} {\bibinfo  {journal}
  {Phys. Rev. Lett.}\ }\textbf {\bibinfo {volume} {118}},\ \bibinfo {pages}
  {027402} (\bibinfo {year} {2017})}\BibitemShut {NoStop}%
\bibitem [{\citenamefont {Latorre}\ and\ \citenamefont
  {Or{\'u}s}(2004)}]{latorre2004adiabatic}%
  \BibitemOpen
  \bibfield  {author} {\bibinfo {author} {\bibfnamefont {Jos{\'e}~Ignacio}\
  \bibnamefont {Latorre}}\ and\ \bibinfo {author} {\bibfnamefont {Rom{\'a}n}\
  \bibnamefont {Or{\'u}s}},\ }\bibfield  {title} {\enquote {\bibinfo {title}
  {Adiabatic quantum computation and quantum phase transitions},}\ }\href@noop
  {} {\bibfield  {journal} {\bibinfo  {journal} {Phys. Rev. A}\ }\textbf
  {\bibinfo {volume} {69}},\ \bibinfo {pages} {062302} (\bibinfo {year}
  {2004})}\BibitemShut {NoStop}%
\bibitem [{\citenamefont {Sch{\"u}tzhold}\ and\ \citenamefont
  {Schaller}(2006)}]{schutzhold2006adiabatic}%
  \BibitemOpen
  \bibfield  {author} {\bibinfo {author} {\bibfnamefont {Ralf}\ \bibnamefont
  {Sch{\"u}tzhold}}\ and\ \bibinfo {author} {\bibfnamefont {Gernot}\
  \bibnamefont {Schaller}},\ }\bibfield  {title} {\enquote {\bibinfo {title}
  {Adiabatic quantum algorithms as quantum phase transitions: First versus
  second order},}\ }\href@noop {} {\bibfield  {journal} {\bibinfo  {journal}
  {Phys. Rev. A}\ }\textbf {\bibinfo {volume} {74}},\ \bibinfo {pages}
  {060304(R)} (\bibinfo {year} {2006})}\BibitemShut {NoStop}%
\bibitem [{\citenamefont {Amin}\ and\ \citenamefont
  {Choi}(2009)}]{amin2009first}%
  \BibitemOpen
  \bibfield  {author} {\bibinfo {author} {\bibfnamefont {Mohammad~HS}\
  \bibnamefont {Amin}}\ and\ \bibinfo {author} {\bibfnamefont {Vicki}\
  \bibnamefont {Choi}},\ }\bibfield  {title} {\enquote {\bibinfo {title}
  {First-order quantum phase transition in adiabatic quantum computation},}\
  }\href@noop {} {\bibfield  {journal} {\bibinfo  {journal} {Phys. Rev. A}\
  }\textbf {\bibinfo {volume} {80}},\ \bibinfo {pages} {062326} (\bibinfo
  {year} {2009})}\BibitemShut {NoStop}%
\bibitem [{\citenamefont {Babin}\ \emph {et~al.}(2013)\citenamefont {Babin},
  \citenamefont {Leforestier},\ and\ \citenamefont
  {Paesani}}]{babin2013development}%
  \BibitemOpen
  \bibfield  {author} {\bibinfo {author} {\bibfnamefont {Volodymyr}\
  \bibnamefont {Babin}}, \bibinfo {author} {\bibfnamefont {Claude}\
  \bibnamefont {Leforestier}}, \ and\ \bibinfo {author} {\bibfnamefont
  {Francesco}\ \bibnamefont {Paesani}},\ }\bibfield  {title} {\enquote
  {\bibinfo {title} {Development of a ?first principles? water potential
  with flexible monomers: Dimer potential energy surface, vrt spectrum, and
  second virial coefficient},}\ }\href@noop {} {\bibfield  {journal} {\bibinfo
  {journal} {J. Chem. Theory Comput.}\ }\textbf {\bibinfo {volume} {9}},\
  \bibinfo {pages} {5395} (\bibinfo {year} {2013})}\BibitemShut {NoStop}%
\bibitem [{\citenamefont {Babin}\ \emph {et~al.}(2014)\citenamefont {Babin},
  \citenamefont {Medders},\ and\ \citenamefont
  {Paesani}}]{babin2014development}%
  \BibitemOpen
  \bibfield  {author} {\bibinfo {author} {\bibfnamefont {Volodymyr}\
  \bibnamefont {Babin}}, \bibinfo {author} {\bibfnamefont {Gregory~R}\
  \bibnamefont {Medders}}, \ and\ \bibinfo {author} {\bibfnamefont {Francesco}\
  \bibnamefont {Paesani}},\ }\bibfield  {title} {\enquote {\bibinfo {title}
  {Development of a ?first principles? water potential with flexible
  monomers. ii: Trimer potential energy surface, third virial coefficient, and
  small clusters},}\ }\href@noop {} {\bibfield  {journal} {\bibinfo  {journal}
  {J. Chem. Theory Comput.}\ }\textbf {\bibinfo {volume} {10}},\ \bibinfo
  {pages} {1599--1607} (\bibinfo {year} {2014})}\BibitemShut {NoStop}%
\bibitem [{\citenamefont {Serwatka}\ and\ \citenamefont
  {Roy}(2022)}]{serwatka2022ground}%
  \BibitemOpen
  \bibfield  {author} {\bibinfo {author} {\bibfnamefont {Tobias}\ \bibnamefont
  {Serwatka}}\ and\ \bibinfo {author} {\bibfnamefont {Pierre-Nicholas}\
  \bibnamefont {Roy}},\ }\bibfield  {title} {\enquote {\bibinfo {title} {Ground
  state of asymmetric tops with dmrg: water in one dimension},}\ }\href@noop {}
  {\bibfield  {journal} {\bibinfo  {journal} {J. Chem. Phys.}\ } (\bibinfo
  {year} {2022})}\BibitemShut {NoStop}%
\bibitem [{sup()}]{supp}%
  \BibitemOpen
  \href@noop {} {}\bibinfo {note} {See Supplemental Material [url] for
  computational methods and scaling analysis, which includes Refs.
  \cite{hall1967pure,white1992density,itensor}}\BibitemShut {NoStop}%
\bibitem [{\citenamefont {Vojta}(2003)}]{vojta2003quantum}%
  \BibitemOpen
  \bibfield  {author} {\bibinfo {author} {\bibfnamefont {Matthias}\
  \bibnamefont {Vojta}},\ }\bibfield  {title} {\enquote {\bibinfo {title}
  {Quantum phase transitions},}\ }\href@noop {} {\bibfield  {journal} {\bibinfo
   {journal} {Rep. Prog. Phys.}\ }\textbf {\bibinfo {volume} {66}},\ \bibinfo
  {pages} {2069} (\bibinfo {year} {2003})}\BibitemShut {NoStop}%
\bibitem [{\citenamefont {Sachdev}\ and\ \citenamefont
  {Keimer}(2011)}]{sachdev2011quantumcrit}%
  \BibitemOpen
  \bibfield  {author} {\bibinfo {author} {\bibfnamefont {Subir}\ \bibnamefont
  {Sachdev}}\ and\ \bibinfo {author} {\bibfnamefont {Bernhard}\ \bibnamefont
  {Keimer}},\ }\bibfield  {title} {\enquote {\bibinfo {title} {Quantum
  criticality},}\ }\href@noop {} {\bibfield  {journal} {\bibinfo  {journal}
  {Phys. Today}\ }\textbf {\bibinfo {volume} {64}},\ \bibinfo {pages} {29}
  (\bibinfo {year} {2011})}\BibitemShut {NoStop}%
\bibitem [{\citenamefont {Fisher}\ and\ \citenamefont
  {Barber}(1972)}]{fisher1972scaling}%
  \BibitemOpen
  \bibfield  {author} {\bibinfo {author} {\bibfnamefont {Michael~E}\
  \bibnamefont {Fisher}}\ and\ \bibinfo {author} {\bibfnamefont {Michael~N}\
  \bibnamefont {Barber}},\ }\bibfield  {title} {\enquote {\bibinfo {title}
  {Scaling theory for finite-size effects in the critical region},}\
  }\href@noop {} {\bibfield  {journal} {\bibinfo  {journal} {Phys. Rev. Lett.}\
  }\textbf {\bibinfo {volume} {28}},\ \bibinfo {pages} {1516} (\bibinfo {year}
  {1972})}\BibitemShut {NoStop}%
\bibitem [{\citenamefont {De~Chiara}\ \emph {et~al.}(2012)\citenamefont
  {De~Chiara}, \citenamefont {Lepori}, \citenamefont {Lewenstein},\ and\
  \citenamefont {Sanpera}}]{de2012entanglement}%
  \BibitemOpen
  \bibfield  {author} {\bibinfo {author} {\bibfnamefont {Gabriele}\
  \bibnamefont {De~Chiara}}, \bibinfo {author} {\bibfnamefont {Luca}\
  \bibnamefont {Lepori}}, \bibinfo {author} {\bibfnamefont {Maciej}\
  \bibnamefont {Lewenstein}}, \ and\ \bibinfo {author} {\bibfnamefont {Anna}\
  \bibnamefont {Sanpera}},\ }\bibfield  {title} {\enquote {\bibinfo {title}
  {Entanglement spectrum, critical exponents, and order parameters in quantum
  spin chains},}\ }\href@noop {} {\bibfield  {journal} {\bibinfo  {journal}
  {Phys. Rev. Lett.}\ }\textbf {\bibinfo {volume} {109}},\ \bibinfo {pages}
  {237208} (\bibinfo {year} {2012})}\BibitemShut {NoStop}%
\bibitem [{\citenamefont {Ibarra-Garc{\'\i}a-Padilla}\ \emph
  {et~al.}(2016)\citenamefont {Ibarra-Garc{\'\i}a-Padilla}, \citenamefont
  {Malanche-Flores},\ and\ \citenamefont
  {Poveda-Cuevas}}]{ibarra2016hobbyhorse}%
  \BibitemOpen
  \bibfield  {author} {\bibinfo {author} {\bibfnamefont {Eduardo}\ \bibnamefont
  {Ibarra-Garc{\'\i}a-Padilla}}, \bibinfo {author} {\bibfnamefont
  {Carlos~Gerardo}\ \bibnamefont {Malanche-Flores}}, \ and\ \bibinfo {author}
  {\bibfnamefont {Freddy~Jackson}\ \bibnamefont {Poveda-Cuevas}},\ }\bibfield
  {title} {\enquote {\bibinfo {title} {The hobbyhorse of magnetic systems: the
  ising model},}\ }\href@noop {} {\bibfield  {journal} {\bibinfo  {journal}
  {Eur. J. Phys.}\ }\textbf {\bibinfo {volume} {37}},\ \bibinfo {pages}
  {065103} (\bibinfo {year} {2016})}\BibitemShut {NoStop}%
\bibitem [{\citenamefont {Calabrese}\ and\ \citenamefont
  {Cardy}(2004)}]{calabrese2004entanglement}%
  \BibitemOpen
  \bibfield  {author} {\bibinfo {author} {\bibfnamefont {Pasquale}\
  \bibnamefont {Calabrese}}\ and\ \bibinfo {author} {\bibfnamefont {John}\
  \bibnamefont {Cardy}},\ }\bibfield  {title} {\enquote {\bibinfo {title}
  {Entanglement entropy and quantum field theory},}\ }\href@noop {} {\bibfield
  {journal} {\bibinfo  {journal} {J. Stat. Mech.: Theor. Exp.}\ }\textbf
  {\bibinfo {volume} {2004}},\ \bibinfo {pages} {P06002} (\bibinfo {year}
  {2004})}\BibitemShut {NoStop}%
\bibitem [{\citenamefont {Calabrese}\ and\ \citenamefont
  {Cardy}(2009)}]{calabrese2009entanglement}%
  \BibitemOpen
  \bibfield  {author} {\bibinfo {author} {\bibfnamefont {Pasquale}\
  \bibnamefont {Calabrese}}\ and\ \bibinfo {author} {\bibfnamefont {John}\
  \bibnamefont {Cardy}},\ }\bibfield  {title} {\enquote {\bibinfo {title}
  {Entanglement entropy and conformal field theory},}\ }\href@noop {}
  {\bibfield  {journal} {\bibinfo  {journal} {J. Phys. A: math. Theor.}\
  }\textbf {\bibinfo {volume} {42}},\ \bibinfo {pages} {504005} (\bibinfo
  {year} {2009})}\BibitemShut {NoStop}%
\bibitem [{\citenamefont {Eisert}\ \emph {et~al.}(2010)\citenamefont {Eisert},
  \citenamefont {Cramer},\ and\ \citenamefont {Plenio}}]{eisert2010colloquium}%
  \BibitemOpen
  \bibfield  {author} {\bibinfo {author} {\bibfnamefont {Jens}\ \bibnamefont
  {Eisert}}, \bibinfo {author} {\bibfnamefont {Marcus}\ \bibnamefont {Cramer}},
  \ and\ \bibinfo {author} {\bibfnamefont {Martin~B}\ \bibnamefont {Plenio}},\
  }\bibfield  {title} {\enquote {\bibinfo {title} {Colloquium: Area laws for
  the entanglement entropy},}\ }\href@noop {} {\bibfield  {journal} {\bibinfo
  {journal} {Rev. Mod. Phys.}\ }\textbf {\bibinfo {volume} {82}},\ \bibinfo
  {pages} {277} (\bibinfo {year} {2010})}\BibitemShut {NoStop}%
\bibitem [{\citenamefont {Katsura}(2012)}]{katsura2012sine}%
  \BibitemOpen
  \bibfield  {author} {\bibinfo {author} {\bibfnamefont {Hosho}\ \bibnamefont
  {Katsura}},\ }\bibfield  {title} {\enquote {\bibinfo {title} {Sine-square
  deformation of solvable spin chains and conformal field theories},}\
  }\href@noop {} {\bibfield  {journal} {\bibinfo  {journal} {J. Phys. A: Math.
  Theor.}\ }\textbf {\bibinfo {volume} {45}},\ \bibinfo {pages} {115003}
  (\bibinfo {year} {2012})}\BibitemShut {NoStop}%
\bibitem [{\citenamefont {Hotta}\ and\ \citenamefont
  {Shibata}(2012)}]{hotta2012grand}%
  \BibitemOpen
  \bibfield  {author} {\bibinfo {author} {\bibfnamefont {Chisa}\ \bibnamefont
  {Hotta}}\ and\ \bibinfo {author} {\bibfnamefont {Naokazu}\ \bibnamefont
  {Shibata}},\ }\bibfield  {title} {\enquote {\bibinfo {title} {Grand canonical
  finite-size numerical approaches: A route to measuring bulk properties in an
  applied field},}\ }\href@noop {} {\bibfield  {journal} {\bibinfo  {journal}
  {Phys. Rev. B}\ }\textbf {\bibinfo {volume} {86}},\ \bibinfo {pages}
  {041108(R)} (\bibinfo {year} {2012})}\BibitemShut {NoStop}%
\bibitem [{\citenamefont {Halverson}\ \emph {et~al.}(2018)\citenamefont
  {Halverson}, \citenamefont {Iouchtchenko},\ and\ \citenamefont
  {Roy}}]{halverson2018quantifying}%
  \BibitemOpen
  \bibfield  {author} {\bibinfo {author} {\bibfnamefont {Tom}\ \bibnamefont
  {Halverson}}, \bibinfo {author} {\bibfnamefont {Dmitri}\ \bibnamefont
  {Iouchtchenko}}, \ and\ \bibinfo {author} {\bibfnamefont {Pierre-Nicholas}\
  \bibnamefont {Roy}},\ }\bibfield  {title} {\enquote {\bibinfo {title}
  {Quantifying entanglement of rotor chains using basis truncation: Application
  to dipolar endofullerene peapods},}\ }\href@noop {} {\bibfield  {journal}
  {\bibinfo  {journal} {The Journal of chemical physics}\ }\textbf {\bibinfo
  {volume} {148}},\ \bibinfo {pages} {074112} (\bibinfo {year}
  {2018})}\BibitemShut {NoStop}%
\bibitem [{\citenamefont {Biskupek}\ \emph {et~al.}(2020)\citenamefont
  {Biskupek}, \citenamefont {Skowron}, \citenamefont {Stoppiello},
  \citenamefont {Rance}, \citenamefont {Alom}, \citenamefont {Fung},
  \citenamefont {Whitby}, \citenamefont {Levitt}, \citenamefont {Ramasse},
  \citenamefont {Kaiser} \emph {et~al.}}]{biskupek2020bond}%
  \BibitemOpen
  \bibfield  {author} {\bibinfo {author} {\bibfnamefont {Johannes}\
  \bibnamefont {Biskupek}}, \bibinfo {author} {\bibfnamefont {Stephen~T}\
  \bibnamefont {Skowron}}, \bibinfo {author} {\bibfnamefont {Craig~T}\
  \bibnamefont {Stoppiello}}, \bibinfo {author} {\bibfnamefont {Graham~A}\
  \bibnamefont {Rance}}, \bibinfo {author} {\bibfnamefont {Shamim}\
  \bibnamefont {Alom}}, \bibinfo {author} {\bibfnamefont {Kayleigh~LY}\
  \bibnamefont {Fung}}, \bibinfo {author} {\bibfnamefont {Richard~J}\
  \bibnamefont {Whitby}}, \bibinfo {author} {\bibfnamefont {Malcolm~H}\
  \bibnamefont {Levitt}}, \bibinfo {author} {\bibfnamefont {Quentin~M}\
  \bibnamefont {Ramasse}}, \bibinfo {author} {\bibfnamefont {Ute}\ \bibnamefont
  {Kaiser}},  \emph {et~al.},\ }\bibfield  {title} {\enquote {\bibinfo {title}
  {Bond dissociation and reactivity of hf and h2o in a nano test tube},}\
  }\href@noop {} {\bibfield  {journal} {\bibinfo  {journal} {ACS nano}\
  }\textbf {\bibinfo {volume} {14}},\ \bibinfo {pages} {11178--11189} (\bibinfo
  {year} {2020})}\BibitemShut {NoStop}%
\bibitem [{\citenamefont {Koch}\ \emph {et~al.}(2019)\citenamefont {Koch},
  \citenamefont {Lemeshko},\ and\ \citenamefont {Sugny}}]{koch2019quantum}%
  \BibitemOpen
  \bibfield  {author} {\bibinfo {author} {\bibfnamefont {Christiane~P}\
  \bibnamefont {Koch}}, \bibinfo {author} {\bibfnamefont {Mikhail}\
  \bibnamefont {Lemeshko}}, \ and\ \bibinfo {author} {\bibfnamefont
  {Dominique}\ \bibnamefont {Sugny}},\ }\bibfield  {title} {\enquote {\bibinfo
  {title} {Quantum control of molecular rotation},}\ }\href@noop {} {\bibfield
  {journal} {\bibinfo  {journal} {Reviews of Modern Physics}\ }\textbf
  {\bibinfo {volume} {91}},\ \bibinfo {pages} {035005} (\bibinfo {year}
  {2019})}\BibitemShut {NoStop}%
\bibitem [{\citenamefont {Albert}\ \emph {et~al.}(2020)\citenamefont {Albert},
  \citenamefont {Covey},\ and\ \citenamefont {Preskill}}]{albert2020robust}%
  \BibitemOpen
  \bibfield  {author} {\bibinfo {author} {\bibfnamefont {Victor~V}\
  \bibnamefont {Albert}}, \bibinfo {author} {\bibfnamefont {Jacob~P}\
  \bibnamefont {Covey}}, \ and\ \bibinfo {author} {\bibfnamefont {John}\
  \bibnamefont {Preskill}},\ }\bibfield  {title} {\enquote {\bibinfo {title}
  {Robust encoding of a qubit in a molecule},}\ }\href@noop {} {\bibfield
  {journal} {\bibinfo  {journal} {Physical Review X}\ }\textbf {\bibinfo
  {volume} {10}},\ \bibinfo {pages} {031050} (\bibinfo {year}
  {2020})}\BibitemShut {NoStop}%
\bibitem [{\citenamefont {Hughes}\ \emph {et~al.}(2020)\citenamefont {Hughes},
  \citenamefont {Frye}, \citenamefont {Sawant}, \citenamefont {Bhole},
  \citenamefont {Jones}, \citenamefont {Cornish}, \citenamefont {Tarbutt},
  \citenamefont {Hutson}, \citenamefont {Jaksch},\ and\ \citenamefont
  {Mur-Petit}}]{hughes2020robust}%
  \BibitemOpen
  \bibfield  {author} {\bibinfo {author} {\bibfnamefont {Michael}\ \bibnamefont
  {Hughes}}, \bibinfo {author} {\bibfnamefont {Matthew~D}\ \bibnamefont
  {Frye}}, \bibinfo {author} {\bibfnamefont {Rahul}\ \bibnamefont {Sawant}},
  \bibinfo {author} {\bibfnamefont {Gaurav}\ \bibnamefont {Bhole}}, \bibinfo
  {author} {\bibfnamefont {Jonathan~A}\ \bibnamefont {Jones}}, \bibinfo
  {author} {\bibfnamefont {Simon~L}\ \bibnamefont {Cornish}}, \bibinfo {author}
  {\bibfnamefont {MR}~\bibnamefont {Tarbutt}}, \bibinfo {author} {\bibfnamefont
  {Jeremy~M}\ \bibnamefont {Hutson}}, \bibinfo {author} {\bibfnamefont
  {Dieter}\ \bibnamefont {Jaksch}}, \ and\ \bibinfo {author} {\bibfnamefont
  {Jordi}\ \bibnamefont {Mur-Petit}},\ }\bibfield  {title} {\enquote {\bibinfo
  {title} {Robust entangling gate for polar molecules using magnetic and
  microwave fields},}\ }\href@noop {} {\bibfield  {journal} {\bibinfo
  {journal} {Physical Review A}\ }\textbf {\bibinfo {volume} {101}},\ \bibinfo
  {pages} {062308} (\bibinfo {year} {2020})}\BibitemShut {NoStop}%
\bibitem [{\citenamefont {Tscherbul}\ \emph {et~al.}(2022)\citenamefont
  {Tscherbul}, \citenamefont {Ye},\ and\ \citenamefont
  {Rey}}]{tscherbul2022robust}%
  \BibitemOpen
  \bibfield  {author} {\bibinfo {author} {\bibfnamefont {Timur~V}\ \bibnamefont
  {Tscherbul}}, \bibinfo {author} {\bibfnamefont {Jun}\ \bibnamefont {Ye}}, \
  and\ \bibinfo {author} {\bibfnamefont {Ana~Maria}\ \bibnamefont {Rey}},\
  }\bibfield  {title} {\enquote {\bibinfo {title} {Robust nuclear spin
  entanglement via dipolar interactions in polar molecules},}\ }\href@noop {}
  {\bibfield  {journal} {\bibinfo  {journal} {arXiv preprint arXiv:2205.00190}\
  } (\bibinfo {year} {2022})}\BibitemShut {NoStop}%
\bibitem [{\citenamefont {Hall}\ and\ \citenamefont
  {Dowling}(1967)}]{hall1967pure}%
  \BibitemOpen
  \bibfield  {author} {\bibinfo {author} {\bibfnamefont {Richard~T}\
  \bibnamefont {Hall}}\ and\ \bibinfo {author} {\bibfnamefont {Jerome~M}\
  \bibnamefont {Dowling}},\ }\bibfield  {title} {\enquote {\bibinfo {title}
  {Pure rotational spectrum of water vapor},}\ }\href@noop {} {\bibfield
  {journal} {\bibinfo  {journal} {J. Chem. Phys.}\ }\textbf {\bibinfo {volume}
  {47}},\ \bibinfo {pages} {2454--2461} (\bibinfo {year} {1967})}\BibitemShut
  {NoStop}%
\bibitem [{\citenamefont {White}(1992)}]{white1992density}%
  \BibitemOpen
  \bibfield  {author} {\bibinfo {author} {\bibfnamefont {Steven~R}\
  \bibnamefont {White}},\ }\bibfield  {title} {\enquote {\bibinfo {title}
  {Density matrix formulation for quantum renormalization groups},}\
  }\href@noop {} {\bibfield  {journal} {\bibinfo  {journal} {Phys. Rev. Lett}\
  }\textbf {\bibinfo {volume} {69}},\ \bibinfo {pages} {2863} (\bibinfo {year}
  {1992})}\BibitemShut {NoStop}%
\bibitem [{\citenamefont {Fishman}\ \emph {et~al.}(2020)\citenamefont
  {Fishman}, \citenamefont {White},\ and\ \citenamefont
  {Stoudenmire}}]{itensor}%
  \BibitemOpen
  \bibfield  {author} {\bibinfo {author} {\bibfnamefont {Matthew}\ \bibnamefont
  {Fishman}}, \bibinfo {author} {\bibfnamefont {Steven~R.}\ \bibnamefont
  {White}}, \ and\ \bibinfo {author} {\bibfnamefont {E.~Miles}\ \bibnamefont
  {Stoudenmire}},\ }\href@noop {} {\enquote {\bibinfo {title} {The
  \mbox{ITensor} software library for tensor network calculations},}\ }
  (\bibinfo {year} {2020}),\ \Eprint {http://arxiv.org/abs/2007.14822}
  {arXiv:2007.14822} \BibitemShut {NoStop}%
\end{thebibliography}
%

\end{document}